\documentclass{ws-ijmpa}

\newcommand{\D}{\ensuremath{D}}
\newcommand{\Dbar}{\ensuremath{\overline{D}}}
\newcommand{\Ds}{\ensuremath{D_{s}}}
\newcommand{\Dsp}{\ensuremath{D_{s}^+}}
\newcommand{\Dsm}{\ensuremath{D_{s}^-}}
\newcommand{\mbc}{\ensuremath{m_{\textrm{BC}}}}

\newcommand{\Dz}{\ensuremath{D^0}}
\newcommand{\Dzb}{\ensuremath{\overline{D}^0}}
\newcommand{\Dp}{\ensuremath{D^+}}
\newcommand{\Dm}{\ensuremath{D^-}}
\newcommand{\Kp}{\ensuremath{K^+}}
\newcommand{\Km}{\ensuremath{K^-}}
\newcommand{\KS}{\ensuremath{K^0_{S}}}
\newcommand{\KL}{\ensuremath{K^0_{L}}}
\newcommand{\KSL}{\ensuremath{K^0_{S,L}}}
\newcommand{\pip}{\ensuremath{\pi^+}}
\newcommand{\pim}{\ensuremath{\pi^-}}
\newcommand{\piz}{\ensuremath{{\pi^0}}}

\newcommand{\invpb}{pb$^{-1}$}

\newcommand{\Kz}{\ensuremath{K^{0}}}
\newcommand{\Kzb}{\ensuremath{\overline{K}^{0}}}

\newcommand{\Ecand}{\ensuremath{E_{\textrm{cand}}}}

\newcommand{\Ecm}{\ensuremath{E_{\textrm{cm}}}}

\begin{document}

\markboth{Peter Onyisi}
{Charm Hadronic Decay Branching Fractions from CLEO-c}

%
\catchline{}{}{}{}{}
%

\title{CHARM HADRONIC DECAY BRANCHING FRACTIONS FROM CLEO-C}

\author{PETER ONYISI\\
(for the CLEO Collaboration)}

\address{Cornell University, Ithaca, New York 14853, USA\\
ponyisi@lepp.cornell.edu}

\maketitle

\begin{history}
\end{history}

\begin{abstract}
The CLEO-c experiment at the CESR $e^+ e^-$ storage ring has collected data
with $\Ecm = 3.77$ GeV and $\Ecm \sim 4.17$ GeV to study the decays of
charmed mesons.  This paper discusses results on the hadronic
branching fractions of the \Dz, \Dp, and \Dsp.  
\end{abstract}

\ccode{PACS numbers: 13.25.Ft, 14.40.Lb }

\section{Introduction}
The branching fractions of all-hadronic decays of the lightest charmed mesons are
of interest for several reasons.  The absolute branching fractions of the \Dz,
\Dp, and \Dsp\ mesons\footnote{Except where stated, mention of a particle or decay
implies the charge conjugate particle or process as well.} serve to normalize
other charm meson decays and decay chains through charm quarks.  Due to the
large branching fraction of $b \to c$, charm branching fractions normalize
many processes in $b$ physics.  Inclusive decay rates of charmed mesons to
final states including easily-identifiable particles can also be used to
disentangle particle content.

Hadronic charm decays are also of intrinsic interest in understanding the
dynamics of the strong force.  Phase shifts between different processes
can be probed by measuring the branching fractions of final states that
interfere or are related by isospin symmetry.

\section{Experimental Methods}
The CLEO-c experiment at the CESR $e^+ e^-$ collider has collected 281 \invpb\
of data at center of mass energy \Ecm\ = 3.77 GeV and approximately 200
\invpb\ with \Ecm\ near 4.17 GeV.  The 3.77 GeV dataset, taken at the peak
of the $\psi(3770)$ resonance, contains approximately 1 million $e^+ e^- \to
\Dz\Dzb$ and 0.8 million $e^+ e^- \to \Dp\Dm$ events.  The 4.17 GeV dataset,
taken at an energy that optimizes the production of $\Ds^{* \pm}\Ds^\mp$
events with cross section $\sim$ 1 nb, is primarily used for studies of \Dsp\
mesons.

For \Dz\ and \Dp\ mesons at 3.77 GeV, and for \Dsp\ mesons at 4.17 GeV, the
production of a charmed meson is always associated with the production of its
antiparticle, as 3.77 GeV and 4.17 GeV are below the $DD\pi$ and $\Ds D K$
thresholds, respectively.  Reconstruction of ``tag'' \Dz, \Dp, or \Dsp\ candidates
thus provides a clean sample of a known number of decays of the antiparticle
and enables powerful methods for measuring absolute branching fractions as
detailed below.

Analyses of the 3.77 GeV data use the kinematic variables $\Delta E$ and \mbc,
defined as\footnote{In this paper, $c=1$, and mass and momentum are
  measured in units of energy.} $\Delta E \equiv \Ecand - E_{\textrm{beam}}$ and
$\mbc \equiv \sqrt{E_{\textrm{beam}}^2 - |\vec{p}_{\textrm{cand}}|^2}$,
where $E_{\textrm{cand}}$, $\vec{p}_{\textrm{cand}}$ are the energy and
momentum of the \D\ candidate.  For equal-mass particles of mass $M$
produced in a two-body process, $\Delta E$ will peak at zero, and \mbc\ will
peak at $M$ with better resolution than the invariant mass.  \D\ candidates
are required to have $\Delta E$ consistent with zero within roughly three
standard deviations.  

For analyses at 4.17 GeV, the \mbc\ variable is used along with the invariant
mass of the \D\ candidate.  At this energy \mbc\ functions as a nearly energy-independent proxy for
momentum, and separates states produced in different two-body processes.  The
\Dsp\ produced from $\Ds^{*+}$ decays are boosted slightly relative to the
nominal two-body momentum; the consequent smearing in \mbc\ is accounted for in the selection
requirements.

\section{Absolute Hadronic Branching Fractions}
Precise measurements of charm meson branching fractions are often made as
branching ratios to easy-to-identify and frequent all-hadronic final states,
especially $\Dz \to \Km\pip$, $\Dp \to \Km\pip\pip$, and $\Dsp\to \phi\pip \to
\Km\Kp\pip$. Uncertainties in these normalizing branching fractions thus
propagate to many other modes.  Additionally, measurements of decay chains
with \D\ mesons in the final state usually involve a normalizing \D\ decay
which must be removed to obtain branching fractions for the intermediate
states.  Reducing the uncertainties in these normalizing modes is a primary
goal of CLEO-c.  Before CLEO-c input, the relative uncertainties in these
branching fractions were about 3\%, 6\%, and 13\% for \Dz, \Dp, and \Dsp,
respectively\cite{pdg04,pdg06}.

To measure the absolute branching fractions CLEO-c uses a technique pioneered
by the the MARK~III collaboration\cite{markiiia,markiiib}.  Single \D\
candidates are reconstructed, regardless of the rest of the event; these are
called ``single tags'' (ST).  Full reconstruction of a \D\Dbar\ pair is also
attempted, and candidates are referred to as ``double tags'' (DT).  Ratios of
ST yields to each other give precise branching ratios, while ratios of DT to
ST yields give information on the absolute branching fractions.  CLEO-c
measures ST yields in three, six, and four decay modes for \Dz, \Dp, and \Dsp\
respectively (separating charge conjugate states), and in 9, 36, and 15 DT
final states for \Dz\Dzb, \Dp\Dm, and \Dsp\Dsm\ respectively (one \Dsp\Dsm\
double tag mode, $\pip\pip\pim/\pim\pim\pip$, is dropped due to continuum
background).  Fits are performed to the observed yields in terms of the total
number of produced \D\ pairs, $N_{\D\Dbar}$, and the branching fraction for
each mode\cite{werner}.

The results here are based on 56 \invpb\ of 3.77 GeV data for \Dz\ and \Dp\
and 75 \invpb\ of $\sim$ 4.17 GeV data for \Dsp.  The \Dz\ and \Dp\ results
are published in Ref.~\refcite{d0dpbfs}; the \Dsp\ results are preliminary.
The branching fractions obtained are listed in Table~\ref{tbl:absbfs}.

\begin{table}[ph]
\tbl{Branching fractions for \Dz\ and \Dp\ (from 56 \invpb) and \Dsp\
  (from 75 \invpb, preliminary).  Uncertainties are
  statistical and systematic, respectively.}
{\begin{tabular}{@{}lc@{}} \toprule
Branching Fraction & Fitted Value (\%)\\
\colrule
$\mathcal{B}(D^0\to K^-\pi^+)$ & 3.91$\pm$ 0.08$\pm$ 0.09\\
$\mathcal{B}(D^0\to K^-\pi^+\pi^0)$ & $14.9\pm 0.3\pm 0.5$ \\
$\mathcal{B}(D^0\to K^-\pi^+\pi^+\pi^-)$ & $8.3\pm 0.2\pm 0.3$ \\
\colrule
$\mathcal{B}(D^+\to K^-\pi^+\pi^+)$ & 9.5$\pm$ 0.2$\pm$ 0.3 \\
$\mathcal{B}(D^+\to K^-\pi^+\pi^+\pi^0)$ & $6.0\pm 0.2\pm 0.2$ \\
$\mathcal{B}(D^+\to \KS\pi^+)$ & $1.55\pm 0.05\pm 0.06$ \\
$\mathcal{B}(D^+\to \KS\pi^+\pi^0)$ & $7.2\pm 0.2\pm 0.4$ \\
$\mathcal{B}(D^+\to \KS\pi^+\pi^+\pi^-)$ & $3.2\pm 0.1\pm 0.2$\\
$\mathcal{B}(D^+\to K^+ K^- \pi^+)$ & $0.97\pm 0.04\pm 0.04$\\
\colrule
$\mathcal{B}(\Dsp\to\KS\Kp)$ & $1.28 ^{+0.13}_{-0.12} \pm 0.07$\\
$\mathcal{B}(\Dsp\to\Km\Kp\pip)$ & $4.54 ^{+0.44}_{-0.42} \pm 0.25$\\
$\mathcal{B}(\Dsp\to\Km\Kp\pip\piz)$ & $4.83 ^{+0.49}_{-0.47} \pm 0.46$ \\
$\mathcal{B}(\Dsp\to\pip\pip\pim)$ & $1.02 ^{+0.11}_{-0.10} \pm 0.05$ \\
\botrule
\end{tabular} \label{tbl:absbfs}}
\end{table}

\begin{figure}[pb]
\centerline{\psfig{file=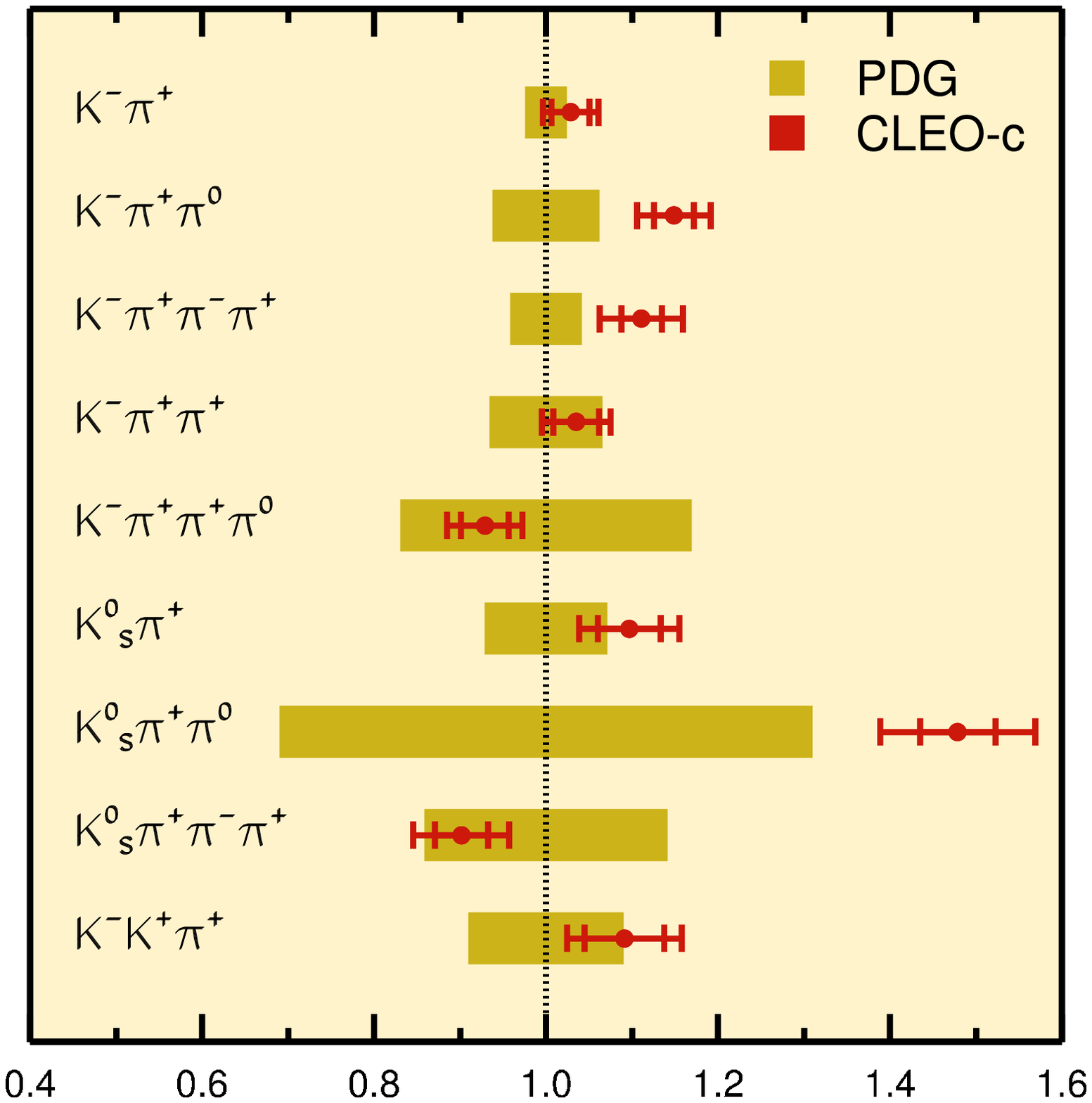,height=4.5cm}
\hskip 1.2cm\psfig{file=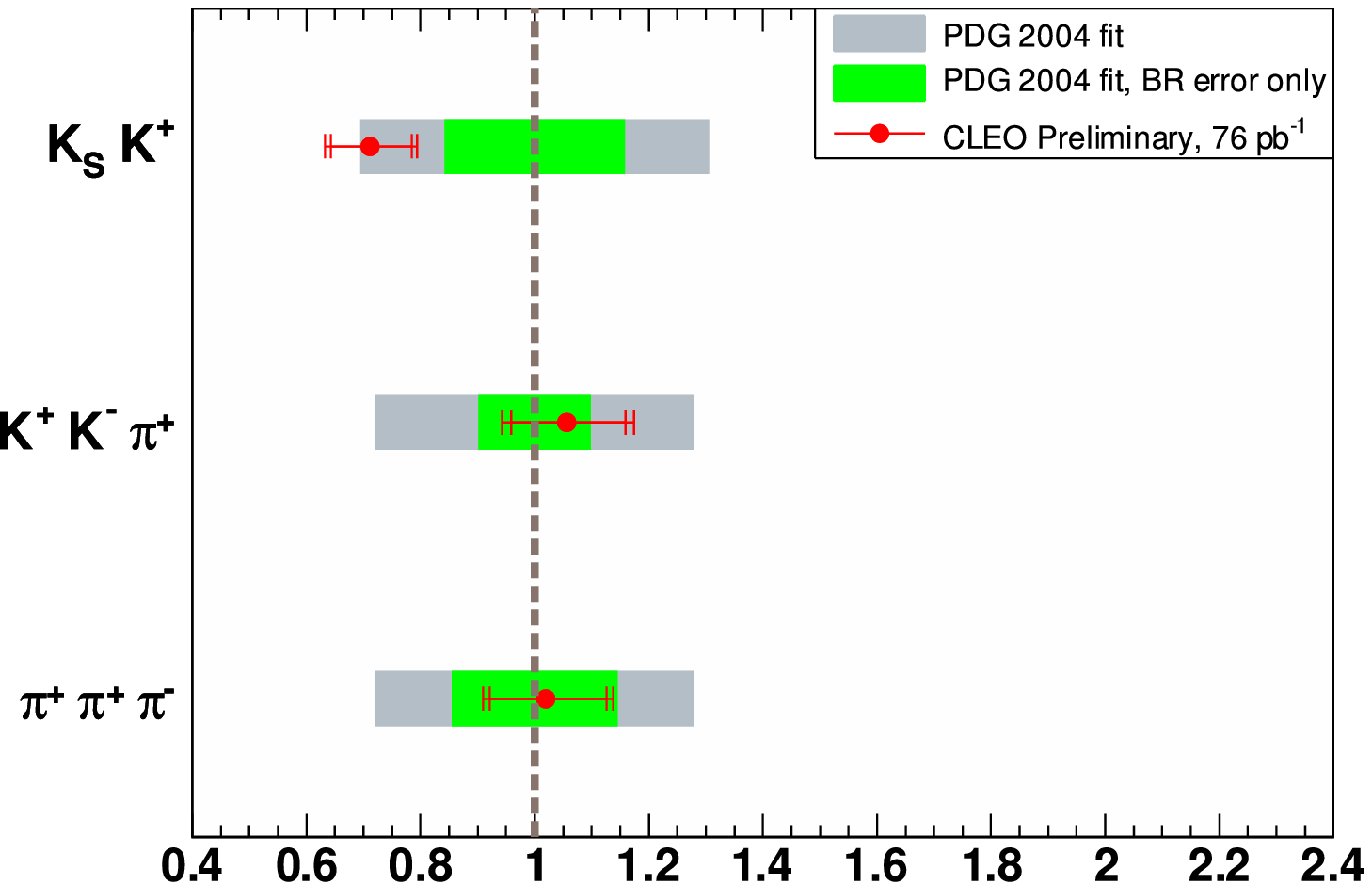,height=4.5cm}}
\vspace*{8pt}
\caption{Ratios of CLEO-c results for absolute \Dz\ and \Dp\ (left) and \Dsp\
  (right)  branching
  fractions to the PDG 2004 fits.  The smaller and larger CLEO-c error bars
  are statistical and systematic uncertainties, respectively.  The bars for
  the \Dsp\ results marked ``PDG 2004 fit, BR error only'' are the uncertainty
  on the branching fraction arising from the branching ratio to
  $\Dsp\to\phi\pip$.  The $\Dsp\to\Km\Kp\pip\piz$ mode is not included as it
  is the first measurement of the inclusive branching fraction to that final state.\label{fig:absbfs}}
\end{figure}

CLEO-c does not report a $\Dsp \to \phi\pip$ branching fraction.  Such events
with $\phi \to \Km\Kp$ are included in the $\Dsp \to \Km\Kp\pip$ branching
fraction.  Previous experiments have evaluated cuts and efficiencies assuming
that all the \Km\Kp\pip\ signal in the $\phi$ mass region is in fact from the
$\phi\pip$ intermediate state.  However there is strong
evidence\cite{e687,focus} for a broad scalar contribution underneath the
$\phi$ which will contribute differently to a ``$\phi\pip$'' signal depending
on the specific selection requirements chosen, possibly varying up to 10\%
between measurements.  Since CLEO-c measurements will soon exceed this level
of precision, the need for better-defined reference branching fractions has
been recognized.  The long-term solution for precision experiments wishing to
use a ``$\phi\pip$'' signal will be to couple the inclusive $\Km\Kp\pip$
branching fraction with Monte Carlo incorporating Dalitz analysis of this
decay.

\section{Cabibbo--Suppressed \Dz\ and \Dp\ Decays}
The branching fractions of the Cabibbo--suppressed decays of \Dz\ and \Dp\ to
multi-pion final states are poorly measured, especially for modes with \piz\
mesons.  Knowledge of these branching fractions allows better understanding
and simulation of backgrounds to other decays, in particular those involving
\KS\ mesons.  Multi-pion decays also give information on intermediate resonant states
(through, {\it e.g.,} $\eta \to \pip\pim\piz$ and $\omega \to \pip\pim\piz$).
Finally, the three decay modes $\Dz \to \pip\pim$, $\Dz \to \piz\piz$, and $\Dp
\to \pip\piz$ form an isospin triangle that permits the extraction of the
phase difference between the $\Delta I=3/2$ and $\Delta
I=1/2$ amplitudes ($A_2$ and $A_0$, respectively).

In this analysis\cite{multipions}, which uses the full 281 \invpb\ dataset, candidates are reconstructed in seven \Dz\ and five \Dp\ all-pion final
states.  Vetoes on the invariant mass of $\pip\pim$ and $\piz\piz$ pairs are
applied to remove contamination from \KS\ decays.  Branching ratios are
obtained relative to $\Dz \to \Km\pip$ and $\Dp \to \Km\pip\pip$.  First
observations are made of four inclusive 
final states and two intermediate states; limits are placed on the inclusive
$\Dz \to \piz\piz\piz$ decay and three intermediate states.  The results are
summarized in Table~\ref{tbl:cabibbosup}.  The ratio of the two isospin amplitudes for
$\D \to \pi\pi$ is found to be:
\[ |A_2/A_0| = 0.420 \pm 0.014 \pm 0.016;\hskip 0.5cm \arg(A_2/A_0) = (86.4 \pm 2.8 \pm
3.3)^\circ \]

\begin{table}[ph]
\tbl{Branching fractions using 281 \invpb\ for \Dz\ and \Dp\ decays into all-pion final
  states. Uncertainties are statistical, systematic, uncertainty on reference
  decay branching fraction, and effects of $CP$ correlation (\Dz\ only).}
{\begin{tabular}{@{}lcc@{}} \toprule
Mode & $\mathcal{B}$ ($10^{-3}$) & PDG 2004 ($10^{-3}$)\\
\colrule
$\Dz\to\pip\pim$ & $1.39 \pm 0.04 \pm 0.04 \pm 0.03 \pm 0.01$ & $1.38 \pm 0.05$\\
$\Dz\to\piz\piz$ & $0.79 \pm 0.05 \pm 0.06 \pm 0.01 \pm 0.01$ & $0.84 \pm 0.22$\\
$\Dz\to\pip\pim\piz$ & $13.2 \pm 0.2 \pm 0.5 \pm 0.2 \pm 0.1$ & $11 \pm 4$\\
$\Dz\to\pip\pip\pim\pim$ & $7.3 \pm 0.1 \pm 0.3 \pm 0.1 \pm 0.1$ & $7.3 \pm 0.5$\\
$\Dz\to\pip\pim\piz\piz$ & $9.9 \pm 0.6 \pm 0.7 \pm 0.2 \pm 0.1$ & \\
$\Dz\to\pip\pip\pim\pim\piz$ & $4.1 \pm 0.5 \pm 0.2 \pm 0.1 \pm 0.0$ & \\
$\Dz\to\omega \pip\pim$ & $1.7 \pm 0.5 \pm 0.2 \pm 0.0 \pm 0.0$ & \\
$\Dz\to\eta \piz$ & $0.62 \pm 0.14 \pm 0.05 \pm 0.01 \pm 0.01$ & \\
$\Dz\to\piz\piz\piz$ & $<$ 0.35 (90\% CL) & \\
$\Dz\to\omega \piz$ & $<$ 0.26 (90\% CL) & \\
$\Dz\to\eta \pip\pim$ & $<$ 1.9 (90\% CL) & \\
\colrule
$\Dp\to\pip\piz$ & $1.25 \pm 0.06 \pm 0.07 \pm 0.04$ & $1.33 \pm 0.22$ \\
$\Dp\to\pip\pip\pim$ & $3.35 \pm 0.10 \pm 0.16\pm 0.12$ & $3.1 \pm 0.4$ \\
$\Dp\to\pip\piz\piz$ & $4.8 \pm 0.3 \pm 0.3 \pm 0.2$ & \\
$\Dp\to\pip\pip\pim\piz$ & $11.6 \pm 0.4 \pm 0.6 \pm 0.4$ & \\
$\Dp\to\pip\pip\pip\pim\pim$ & $1.60 \pm 0.18 \pm 0.16 \pm 0.06$ & $1.73 \pm 0.23$ \\
$\Dp\to\eta\pip$ & $3.61 \pm 0.25 \pm 0.23 \pm 0.12$ & $3.0 \pm 0.6$\\
$\Dp\to\omega \pip$ & $<$ 0.34 (90\% CL) & \\
\botrule
\end{tabular} \label{tbl:cabibbosup}}
\end{table}

\section{Decays of \Dp\ to $\KS \pip$ and $\KL \pip$}
The Cabibbo--favored (CF) decay of a \Dz\ or \Dp\ meson to a state with a
neutral kaon includes a \Kzb.  However there are also
doubly-Cabibbo--suppressed (DCS) decays where the final state instead has a
\Kz.  The
$\D \to \Kz X$ amplitude will interfere with the $\D \to \Kzb X$ amplitude
with the sign of the interference being opposite for the observed states \KS\ and \KL.
The branching fractions of \D\ mesons to \KS\ and \KL\
should thus in general be different, with an asymmetry of the order of $\tan^2 \theta_C$
(enhanced over the DCS rate because this is an interference
effect)\cite{bigi-yamamoto}.  This will be modified by the unknown phase
between the CF and DCS amplitudes in each mode.

CLEO-c obtains absolute branching fractions for the process $\Dp \to \KSL
\pip$ from the full 281 \invpb\ sample (``\KSL'' represents the sum of \KS\
and \KL\ contributions to a neutral kaon branching fraction).  Six decay modes
are used to find tag $D^\pm$ candidates.  A pion of the opposite charge,
putatively from the $\D^\mp$ decay, is then searched for, and the missing mass
squared of the $D^\pm \pi^\mp$ system is computed.  A clear peak at the kaon
mass squared is seen.  Extra peaking contributions are seen from two-body
modes with charged pions and muons in the final state, and small smooth
backgrounds from three-body modes and combinatoric background are also
accounted for,
with shapes obtained from Monte Carlo simulations.

\begin{figure}[pt]
\centerline{\psfig{file=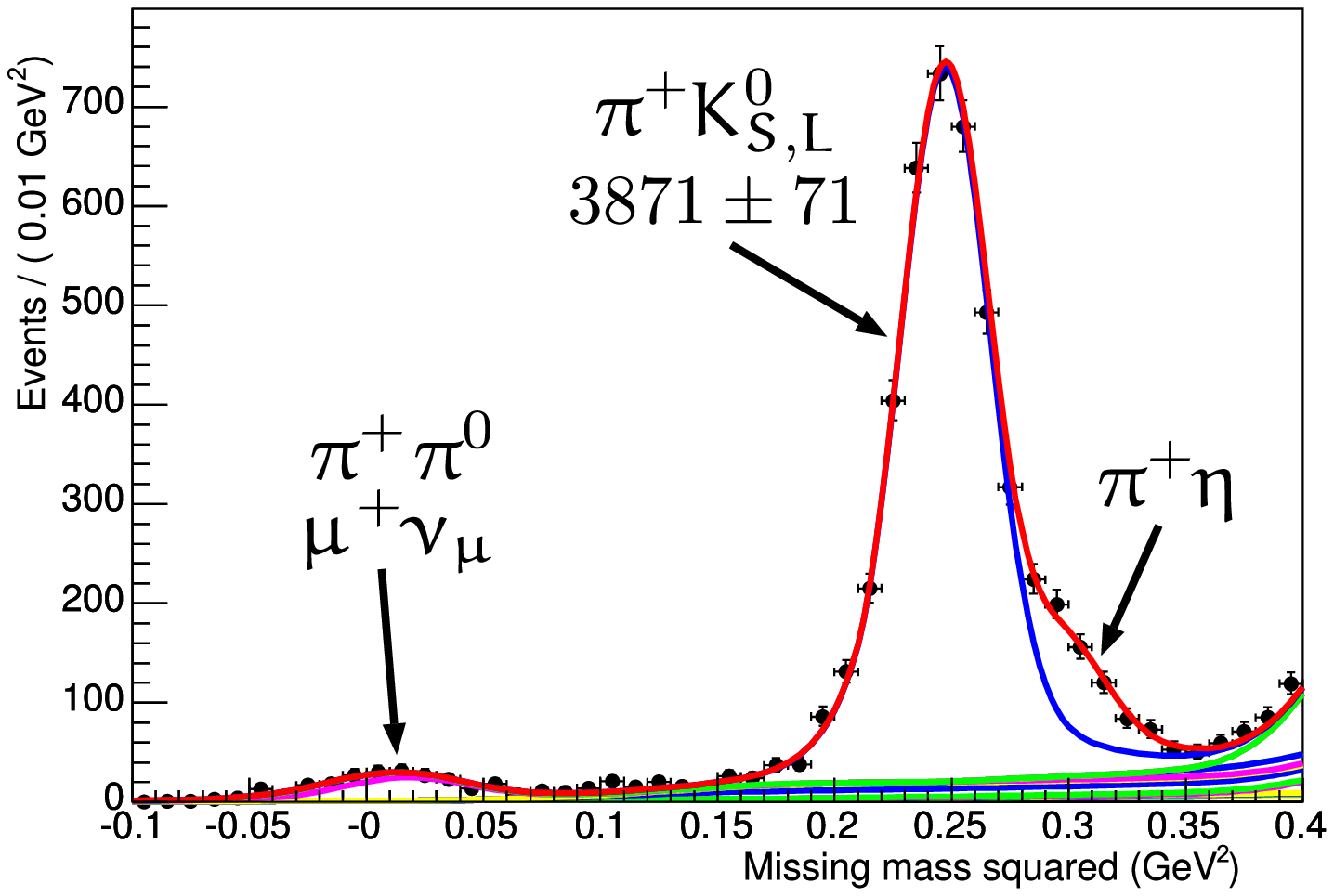,height=6cm}}
\vspace*{8pt}
\caption{Missing mass squared for events in the $\Dp \to \KSL\pip$ analysis. \label{fig:kslpi}}
\end{figure}

A measurement of $\mathcal{B}(\Dp \to \KSL \pip)$ can be combined with the
directly-measured $\mathcal{B}(\Dp \to \KS \pip)$ from Table~\ref{tbl:absbfs}
to obtain an asymmetry.  The analysis yields the following preliminary results:
\[ \mathcal{B}(\Dp \to \KSL \pip) = (3.055 \pm 0.057 \pm
0.158)\% \]
\[\frac{\mathcal{B}(\Dp \to K_L \pip) - \mathcal{B}(\Dp \to K_S \pip)}{\mathcal{B}(\Dp \to \KSL \pip)} = -0.01 \pm 0.04 \pm 0.07\]
\[\mathcal{B}(\Dp \to \eta \pip) = (0.391 \pm 0.031 \pm 0.033)\%\]
No significant asymmetry is observed in this decay mode.  This could result
from the phase between the CF and DCS amplitudes being $\approx 90^\circ$.
The branching fraction for $\Dp \to \eta \pip$ agrees with that obtained from
the multi-pion analysis.

\section{Inclusive Decays of Charmed Mesons to $\eta$, $\eta'$, and $\phi$}
Because a Cabibbo--favored decay of a \Dsp\ meson has an $s\bar s$ quark pair,
while Cabibbo--favored \Dz\ and \Dp\ decays generally do not, one expects
that mesons with large $s\bar s$ content will be produced more often in \Dsp\
decay than in \Dz\ and \Dp\ decays.  This difference can be used, for example,
to estimate the rate of $B_s$ production at the $\Upsilon(5S)$ resonance by
exploiting the large $B_s \to D_s X$ branching fraction.

CLEO-c has preliminary measurements of inclusive rates of the production of
$\eta$, $\eta'$, and $\phi$ mesons in \Dz, \Dp, and \Dsp\ decays using the full
281 \invpb\ dataset at 3.77 GeV and 71 \invpb\ of 4.17 GeV data.  \Dz, \Dp, and
\Dsp\ candidates are reconstructed as tags, and the remaining showers and
tracks are used to reconstruct mesons in the $\eta \to \gamma\gamma$, $\eta'
\to \pip\pim\eta \to \pip\pim\gamma\gamma$, and $\phi \to \Km\Kp$ decay
modes.  Sidebands in $\Delta E$ (for \Dz\ and \Dp) and \mbc\ (for \Dsp) are
used to subtract peaking combinatoric backgrounds.

The results are listed in Table~\ref{tbl:phietaetap}.  The \Dsp\ branching
fractions to these mesons are significantly larger than those for \Dz\ and
\Dp, as expected.

\begin{table}[ph]
\tbl{Preliminary branching fractions for \Dz, \Dp, and \Dsp\ decays into inclusive final
  states with $\eta$, $\eta'$, and $\phi$ mesons.  Results use 281 \invpb\ of
  data for \Dz\ and \Dp\ and 71 \invpb\ for \Dsp.}
{\begin{tabular}{@{}cccc@{}} \toprule
Meson & $\mathcal{B}(\D \to \phi X)$ (\%) & $\mathcal{B}(\D \to \eta X)$ (\%)
  & $\mathcal{B}(\D \to \eta'
X)$ (\%)\\
\hline
\Dz & 1.0 $\pm$ 0.1 $\pm$ 0.1 &
      9.4 $\pm$ 0.4 $\pm$ 0.6 &
      2.6 $\pm$ 0.2 $\pm$ 0.2 \\
\Dp & 1.1 $\pm$ 0.1 $\pm$ 0.2 &
      5.7 $\pm$ 0.5 $\pm$ 0.5 &
      1.0 $\pm$ 0.2 $\pm$ 0.1 \\
\Dsp & 15.1 $\pm$ 2.1 $\pm$ 1.5 &
       32.0 $\pm$ 5.6 $\pm$ 4.7 &
       11.9 $\pm$ 3.3 $\pm$ 1.2 \\
\botrule
\end{tabular} \label{tbl:phietaetap}}
\end{table}

\section{Summary}
The CLEO-c experiment has recorded 281 \invpb\ of $e^+ e^-$ collisions at 3.77
GeV and 200 \invpb\ near 4.17 GeV (of the latter, $\sim$ 75 \invpb\ is used
for the results presented here).  Using these data branching fractions for
many decay modes of the \Dz, \Dp, and \Dsp\ mesons have been obtained with
precision comparable to or exceeding the world average of previous
measurements, and the absolute scale of these decays has been
established using the unique kinematics of threshold production.  CLEO-c
intends to increase these datasets to roughly three times their current size.

\section*{Acknowledgments}
This material is based upon work supported by the National Science 
Foundation under grant PHY-0202078 and an NSF
Graduate Research Fellowship.  I thank A.~Ryd, S.~Stroiney, R.~Sia, S.~Stone,
and S.~Blusk for helpful discussions.

\end{document}